\documentclass[a4paper]{jpconf}
\usepackage{graphicx}
\usepackage{txfonts}
\usepackage{hyperref}
\hypersetup{
	unicode=true,            
	pdftoolbar=false,       
	pdfmenubar=false,        
	pdffitwindow=false,      
	pdfstartview={FitH},     
	pdftitle={},             
	pdfauthor={},            
	pdfsubject={},           
	pdfcreator={},           
	pdfproducer={},          
	pdfkeywords={},          
	pdfnewwindow=true,       
	colorlinks=true,         
	linkcolor=blue,          
	citecolor=magenta,       
	filecolor=green,         
	urlcolor=blue            
}

\newcommand{\To}{\longrightarrow}

\newcommand \ket[1]{|\,{#1}\,\rangle}

\begin{document}
\title{Classical conformal blocks, Coulomb gas integrals,\\ and quantum integrable models\footnote{Based on a talk given at the XII. International Symposium on Quantum Theory and Symmetries, July 24-28 , 2023, Prague, Czech Republic.}}

\author{Marcin R.~Pi\c{a}tek}

\address{Institute of Physics, University of Szczecin,
	     Wielkopolska 15, 70-451 Szczecin, Poland}

\ead{marcin.piatek@usz.edu.pl}

\begin{abstract}
In this paper, we recall Richardson's solution of the reduced BCS model, its relationship with the Gaudin model, and the known implementation of these models in conformal field theory. The CFT techniques applied here are based on the use of the free field realization, or more precisely, on the calculation of saddle-point values of Coulomb gas integrals representing certain (perturbed) WZW conformal blocks. We identify the saddle-point limit as the classical limit of conformal blocks. We show that this observation implies a new method for calculating classical conformal blocks and can be further used in the study of quantum integrable models.
\end{abstract}

\section{Introduction}
In this note, we discuss the Richardson and Gaudin quantum integrable models and their implementation in conformal field theory (CFT). We point out that the latter is related to the classical limit of conformal blocks. Exploring this relationship in depth may lead to new methods for analyzing quantum many-body systems, on the one hand, and for obtaining novel results concerning conformal blocks, on the other hand. We give examples of these possibilities.

Conformal blocks ${\cal F}(\left\lbrace\Delta_i\right\rbrace_{i=1}^{n},
\lbrace{\Delta}_p\rbrace_{p=1}^{3g-3+n},c\,|\,\cdot\,)$
represent holomorphic contributions to physical correlation functions. 
Although they are fully determined by conformal symmetry, they are not known in a closed form except for a few examples.
These functions depend on the central charge $c$ of the Virasoro algebra, external conformal weights 
$\left\lbrace\Delta_i\right\rbrace_{i=1}^{n}$, conformal weights $\lbrace{\Delta}_p\rbrace_{p=1}^{3g-3+n}$ 
in the intermediate channels, vertex operators locations, and modular parameters in case of conformal field theories living on surfaces with genus $g>0$.
Lately, a central issue concerning conformal blocks is the problem of calculating their classical limit: 
\begin{equation}\label{Cinf}
c\To\infty \quad\Longleftrightarrow\quad  b\To 0\quad{\rm or}\quad b\To\infty \quad{\rm for}\;\;c=1+6\left(b+\frac{1}{b}\right)^2.
\end{equation}
Based on concrete examples one can conjecture how conformal blocks behave in the classical limit.
If all the weights are heavy, i.e., $(\Delta_i,\Delta_p)=b^2(\delta_i,\delta_p)$ and $\delta_i,\delta_p={\cal O}(b^0)$
then in the limit (\ref{Cinf}) blocks exponentiate to functions known as Zamolodchikovs’ classical conformal blocks
\cite{ZZ}:\footnote{Analogously for $b\To 0$ and $(\Delta_i,\Delta_p)=b^{-2}(\delta_i,\delta_p)$.}
\begin{equation}\label{cla1}
	{\cal F}\!\left(\lbrace\Delta_i\rbrace,
	\lbrace\Delta_p\rbrace, c\,|\,\cdot\,\right)
	\;\stackrel{b\to\infty}{\sim}
	\;{\rm e}^{b^2 f\left(\lbrace\delta_i\rbrace,\lbrace\delta_p\rbrace|\,\cdot\,\right)}.
\end{equation}
If the external weights are heavy and light 
($\lim_{b\to\infty}b^{-2}\Delta^{\sf light}_{k}=0$)
then in the classical limit conformal blocks decompose into a product of the ``light contribution'' $\psi_{\sf light}\!\left(\cdot\right)$ and the exponent of the classical block:\footnote{Analogously for $b\To 0$ and $\lim_{b\to 0}b^{2}\Delta^{\sf light}_{k}=0$.}
\begin{equation}\label{cla2}
	{\cal F}\!\left(\lbrace\Delta_l\rbrace\cup\lbrace\Delta_{k}^{\sf light}\rbrace,
	\lbrace\Delta_p\rbrace, c\,|\,\cdot\,\right)
	\;\stackrel{b\to \infty}{\sim}\psi_{\sf light}\!\left(\cdot\right)\,
	{\rm e}^{b^2 f\left(\lbrace\delta_l\rbrace,\lbrace\delta_p\rbrace|\,\cdot\,\right)}.
\end{equation}
If all the weights are fixed then in the limit $c\To\infty$ conformal blocks reduce to the so-called global blocks, i.e., contributions to the correlation functions
from representations of the ${\rm sl}(2,\mathbf{C})$ algebra.

It turns out that semiclassical asymptotics of conformal blocks have fascinating mathematical and physical applications. Monodromy problems, uniformization, hyperbolic geometry, string field theory, Bethe/gauge and AGT correspondences, entanglement, quantum chaos, thermalization, AdS$_{3}$/CFT$_{2}$ holography,  and perturbation theory of black holes are just some of the topics in which the classical limit of conformal blocks is used. The present article discusses yet one more such area of application, i.e., in the field of quantum integrable systems. This exposition is partially based on the work \cite{PNPP}.

\section{Models of Richardson and Gaudin}
The Richardson model, also known as the reduced BCS model, is defined by the Hamiltonian,
\begin{equation}\label{HrBCS}
	\hat{\sf H}_{\sf rBCS}=
	\sum\limits_{j,\sigma=\pm}\varepsilon_{j\sigma}c_{j\sigma}^{\dagger}c_{j\sigma}
	-gd\sum\limits_{j,j'}c_{j+}^{\dagger}c_{j-}^{\dagger}c_{j'-}c_{j'+}
\end{equation}
which consists of a kinetic term and an interaction term describing the attraction between Cooper pairs.
Here, $c_{j\sigma}^{\dagger}$, $c_{j\sigma}$ are the fermion 
creation and annihilation operators in time-reversed states 
$\ket{j,\pm}$ with energies $\varepsilon_{j\pm}$, $j=1,\ldots,\Omega$. 
The Hamiltonian (\ref{HrBCS}) is a simplified version of the Bardeen-Cooper-Schrieffer (BCS) Hamiltonian, where all 
couplings have been set equal to a single one, namely $g$. The constant $d$ is a mean level spacing.
$\hat{\sf H}_{\sf rBCS}$ can be written in terms of the ``hard-core'' boson operators 
$b_{j}^{\dagger}=c_{j+}^{\dagger}c_{j-}^{\dagger}$, $b_{j}=c_{j-}c_{j+}$
which create, annihilate fermion pairs, respectively and obey the following 
commutation rules
$[b_{j},b_{j'}^{\dagger}]=\delta_{j,j'}(1-2\hat{\sf N}_{j})$,
$\hat{\sf N}_{j}=b_{j}^{\dagger}\,b_{j}$.
The Hamiltonian (\ref{HrBCS}) rewritten in terms of these operators reads as follows:
\begin{equation}\label{HrBCS2}
\hat{\sf H}_{\sf rBCS}=\sum\limits_{j}2\varepsilon_{j}b_{j}^{\dagger}b_{j}
-gd\sum\limits_{j,j'}b_{j}^{\dagger}b_{j'}.
\end{equation}
As above, the sums are taken over a set $\Omega$ of doubly degenerate energy levels $\varepsilon_{j\pm}$.
In the 1960s Richardson exactly solved an eigenvalue problem for (\ref{HrBCS2})
through the Bethe ansatz \cite{R1,RS}. 
Richardson proposed an ansatz for an exact eigenstate, namely,
$\ket{N}=\prod_{\nu=1}^{N}B^{\dagger}_{\nu}\ket{0}$,
where the pair operators
$B^{\dagger}_{\nu}=\sum_{j=1}^{\Omega}b^{\dagger}_{j}/(2\varepsilon_j-u_{\nu})$
have the form appropriate to the solution of the one-pair problem.
The quantities $u_{\nu}$ are pair energies. They are understood as auxiliary parameters which 
are then chosen to fulfill the eigenvalue equation
$\hat{\sf H}_{\sf rBCS}\,\ket{N}={\sf E}_{\sf rBCS}(N)\,\ket{N}$,
where ${\sf E}_{\sf rBCS}(N)\!=\!\sum_{\nu=1}^{N}u_{\nu}$.
The state  $\ket{N}$ is an eigenstate of $\hat{\sf H}_{\sf rBCS}$ if the $N$ pair energies $u_{\nu}$ 
are, complex in general, solutions of the (Bethe ansatz) equations:
\begin{equation}\label{Req}
\frac{1}{gd}+\sum\limits_{i=1}^{\Omega}\frac{1}{u_\nu-z_i}
=\sum\limits_{\mu\neq\nu}^{N}\frac{2}{u_\nu-u_\mu}\;\;\;\;
{\rm for}\;\;\nu=1,\ldots,N,\quad z_i=2\varepsilon_{i}.
\end{equation}

There is a connection between the Richardson model and a class of integrable
spin models obtained by Gaudin.
Indeed, in 1976 Gaudin proposed the so-called rational, trigonometric and elliptic integrable 
models based on sets of certain commuting Hamiltonians \cite{Gaudin1,Gaudin2}.
The simplest model in this family is the rational model defined by a collection of the following Hamiltonians:
\begin{equation}\label{Gaudin}
\hat{\sf H}_{{\sf G}, i} = \sum\limits_{j\neq i}^{\Omega}
\frac{1}{\varepsilon_i-\varepsilon_j}
\left[{\sf t}_{i}^{0}{\sf t}_{j}^{0}
+\frac{1}{2}\left({\sf t}_{i}^{+}{\sf t}_{j}^{-}+{\sf t}_{i}^{-}{\sf t}_{j}^{+}\right)\right]
=:\sum\limits_{j\neq i}^{\Omega}
\frac{{\bf t}_i\cdot{\bf t}_j}{\varepsilon_i-\varepsilon_j}.
\end{equation}
Each separate spin corresponds to a spin-$\frac{1}{2}$ realization of the ${\rm su}(2)$ algebra generated by 
${\sf t}^{0}$, ${\sf t}^{+}$, ${\sf t}^{-}$.
The spin-$\frac{1}{2}$ generators can be written in terms of
the hard-core boson operators: ${\sf t}_{j}^{+}=b_{j}^{+}$, ${\sf t}_{j}^{-}=b_{j}$,
${\sf t}_{j}^{0}=\frac{1}{2}-\hat{\sf N}_j$.
Therefore, $\hat{\sf H}_{{\sf G}, i}$
can be diagonalized by means of the Richardson method. As before the energy eigenvalue is given by
${\sf E}_{{\sf G},i}(N)=\sum_{\nu=1}^{N}u_{\nu}$, but this time the parameters $u_{\nu}$
satisfy equations which are nothing but the Richardson equations (\ref{Req}) in the limit $g\To\infty$.

In 1997 Cambiagio, Rivas and Saraceno (CRS) uncovered 
\cite{Cambiaggio:1997vz} that conserved charges of the reduced BCS model
are given in terms of the rational Gaudin Hamiltonians, i.e., 
$\hat{\sf R}_i=-{\sf t}_{i}^{0}-gd\,\hat{\sf H}_{{\sf G},i}$.
The quantum integrals of motion $\hat{\sf R}_i$ itself can be seen as a set of commuting Hamiltonians. This is a
famous Gaudin model of magnets also known as the central spin model.\footnote{
Actually, it describes a central spin at position ``$0$'' which is coupled to bath spins through long-range interactions,
$\hat{\sf H}={\rm B}{\sf s}_{0}^{z}+2\sum_{j=1}^{\Omega}
({\bf s}_0\cdot{\bf s}_j)/(\varepsilon_{0}-\varepsilon_{j})$.
Here, $\varepsilon_{0}=0$  and $\varepsilon_{j}$ are energy levels of the Richardson-rBCS model. 
The magnetic field has been chosen as  ${\rm B}=-2/g$ and $d=1$.}
Knowing $\hat{\sf R}_i$ one can express $\hat{\sf H}_{\sf rBCS}$ 
in terms of these quantum integrals of motion. As a result one gets:
\begin{equation}\label{HR}
\hat{\sf H}_{\sf rBCS}=\hat{\sf H}_{\sf XY} 
+ \sum\limits_{j=1}^{\Omega}\varepsilon_{j}+gd\left(\frac{1}{2}\Omega-N\right),\;\;\;\;
\hat{\sf H}_{\sf XY}=\sum_{j=1}^{\Omega}2\varepsilon_{j}\hat{\sf R}_{j}
+gd\Big(\sum_{j=1}^{\Omega}\hat{\sf R}_j\Big)^2\nonumber-\frac{3}{4}gd\,\Omega.
\end{equation}
Eq.~(\ref{HR}) opens a possibility to calculate eigenvalues of $\hat{\sf R}_i$ 
by applying Richardson’s solution of the spectral problem for $\hat{\sf H}_{\sf rBCS}$.
However, the eigenvalues of CRS operators have been 
computed in a different way. More specifically, in 2000 Sierra found \cite{Sierra:1999mp} 
closed expression for  them, i.e.,
\begin{equation}\label{Li}
\lambda_i=\frac{gd}{2}\frac{\partial U({\bf z},{\bf u}^{c})}{\partial z_i}\Big|_{z_i=2\varepsilon_i}
=-\frac{1}{2}+gd\left(\sum\limits_{\nu=1}^{N}\frac{1}{2\varepsilon_{i}-u_{\nu}^{\sf c}}
-\frac{1}{4}\sum\limits_{j\neq i}^{\Omega}\frac{1}{\varepsilon_{i}-\varepsilon_{j}}\right),
\end{equation}
using methods of CFT.
The quantity $U({\bf z},{\bf u}^{\sf c})$ named ``Coulomb energy'' in \cite{Sierra:1999mp} is the 
critical value of the ``potential'':
\begin{eqnarray}\label{U}
U({\bf z},{\bf u})&=&
-\sum\limits_{i<j}^{\Omega}\log(z_i-z_j)
-4\sum\limits_{\nu<\mu}^{N}\log(u_\nu-u_\mu)\nonumber\\
&+&
2\sum\limits_{i=1}^{\Omega}\sum\limits_{\nu=1}^{N}\log(z_i-u_\nu)
+\frac{1}{gd}\left(-\sum\limits_{i=1}^{\Omega}z_i+2\sum\limits_{\nu=1}^{N}u_{\nu}\right).
\end{eqnarray}
Here, ${\bf u}^{c}=(u_{1}^{c},\ldots,u_{N}^{c})$ is a solution of 
the conditions $\partial U({\bf z},{\bf u})/\partial u_{\nu}=0$, $\nu=1,\ldots,N$
which are nothing but the Richardson equations (\ref{Req}).
To solve eigenproblems for the Richardson model conserved charges Sierra has shown in \cite{Sierra:1999mp} 
that the Knizhnik-Zamolodchikov (KZ) equation obeyed by the $\widehat{{\rm su}(2)}_{k}$ WZW block, i.e.,
\begin{equation}
\left(\kappa\partial_{z_i}-\sum_{j\neq i}^{\Omega+1}
({\bf t}_i\cdot{\bf t}_j)/(z_i-z_j)\right)\psi^{\sf WZW}(z_1,\ldots,z_{\Omega+1})=0,
\quad\quad
\kappa=(k+2)/2
\end{equation}
is completely equivalent to the following:
\begin{equation}\label{EigenR}
(2gd)^{-1}\hat{\sf R}_{i}\psi=-\kappa\partial_{z_i}\psi,
\quad\quad
\psi^{\sf WZW}=\exp\left[(2gd\kappa)^{-1}\hat{\sf H}_{\sf XY}\right]\psi.
\end{equation}
Here, $\psi\!\!=\!\!\psi^{\rm CG}_{\bf m}({\bf z})$ is certain perturbed WZW conformal block in the free field 
(Coulomb gas) representation.
More precisely, $\psi^{\rm CG}_{\bf m}({\bf z})$ consists of (i) the $\widehat{\rm su(2)}_{k}$ WZW chiral primary fields
$\Phi_{m}^{j}(z)=\left(\gamma(z)\right)^{j-m}{\sf V}_{\alpha}(z)$ 
built out of the $\gamma$-field of the $\beta\gamma$-system and Virasoro chiral vertex operators 
${\sf V}_{\alpha}(z)$ represented as normal ordered exponentials  
with conformal weights  $\Delta_{\alpha}=\alpha(\alpha-2\alpha_0)=j(j+1)/(k+2)$;\footnote{Here, 
$\alpha=(k+2)^{-\frac{1}{2}}j=-2\alpha_0 j$.} (ii) WZW screening charges; (iii) an additional operator 
${\sf V}_{gd}$ which breaks conformal invariance. Within this realization to every energy level 
$z_i=2\varepsilon_i$ corresponds the field $\Phi_{m_i}^{j}(z_i)$
with the spin $j=\frac{1}{2}$ and the ``third component'' $m_i=\frac{1}{2}$
(or $m_i=-\frac{1}{2}$) if the corresponding energy level is empty (or occupied) by a pair of fermions.
Integration variables $u_{\nu}$ in screening operators are the Richardson parameters.
The operator ${\sf V}_{gd}$ implements the coupling $gd$ and is a source of the term $\frac{1}{gd}$ in 
the Richardson equations (\ref{Req}). 
After ordering, $\psi^{\rm CG}_{\bf m}({\bf z})$ has a structure of a multidimensional contour integral,
\begin{equation}\label{RCG}
\psi^{\rm CG}_{\bf m}({\bf z})=\oint\limits_{C_1}{\rm d}u_{1}\ldots
\oint\limits_{C_N}{\rm d}u_{N}\,
\psi^{\beta\gamma}_{\bf m}({\bf z},{\bf u}){\rm e}^{-\alpha_{0}^{2}U({\bf z},{\bf u})},
\end{equation}
and in the limit $\alpha_{0}\To\infty$ $\Leftrightarrow$ $k\To -2$ $\Leftrightarrow$ $\kappa\To 0$
can be calculated using the saddle point method. 
The stationary solutions of $U({\bf z},{\bf u})$ are then given by the solutions of the Richardson equations.
After all one gets 
$\psi^{\rm CG}_{\bf m}({\bf z})\sim\psi^{\sf R}{\rm e}^{-\alpha_{0}^{2}U({\bf z},{\bf u}^{\sf c})}$
for $\alpha_{0}\To\infty$, where $\psi^{\sf R}$ is the Richardson wave function.
Using this asymptotic limit to the equation (\ref{EigenR}) one obtains
$\hat{\sf R}_{i}\psi^{\sf R}=\lambda_{i}\psi^{\sf R}$ in the limit $\kappa\To 0$, 
where $\lambda_{i}$ are given by (\ref{Li}).

As a final remark in this section let us note that the Coulomb energy $U({\bf z},{\bf u}^{\sf c})$ 
and eigenvalues $\lambda_i$ depend on the Richardson parameters 
${\bf u}^{\sf c}=(u_{1}^{\sf c},\ldots,u_{N}^{\sf c})$.
It would be nice to have techniques that allow to calculate functions 
such as $U({\bf z},{\bf u}^{\sf c})$ without need to solve the Bethe ansatz equations.
In our opinion, it is possible to develop such a method.

\section{Virasoro analogues of the Coulomb energy}
As an example of the last statement in the previous section let us consider first 
the Coulomb gas representation of some spherical four-point block, namely, 
\begin{eqnarray*}\label{ZDF}
{\sf Z}(\,\cdot\,|{\bf z}_f)&=&\left\langle 
:{\rm e}^{\hat\alpha_{1}\phi(0)}::{\rm e}^{\hat\alpha_{2}\phi(x)}:
:{\rm e}^{\hat\alpha_{3}\phi(1)}::{\rm e}^{\hat\alpha_{4}\phi(\infty)}:
\left[\int\limits_{0}^{x}:\!{\rm e}^{b \phi(u)}\!:{\rm d}u\right]^{N_1}
\left[\int\limits_{0}^{1}:\!{\rm e}^{b \phi(u)}\!:{\rm d}u\right]^{N_2}
\right\rangle\nonumber\\
&&\hspace{-60pt}=
x^\frac{\alpha_1\alpha_2}{2\beta}(1-x)^\frac{\alpha_2\alpha_3}{2\beta}
\prod\limits_{\mu=1}^{N_1}\int\limits_{0}^{x}{\rm d}u_\mu
\prod\limits_{\mu=N_{1}+1}^{N_{1}+N_{2}}\int\limits_{0}^{1}{\rm d}u_\mu
\prod\limits_{\mu<\nu}\left(u_\nu - u_\mu\right)^{2\beta}
\prod\limits_{\mu}
u_{\mu}^{\alpha_1}\left(u_\mu-x\right)^{\alpha_2}\left(u_\mu-1\right)^{\alpha_3},
\end{eqnarray*}
where ${\bf z}_f:=(0,x,1,\infty)$.
It was not clear for a long time how to choose integration contours to get 
an integral representation of the four-point block
consistent with historically first Belavin-Polyakov-Zamolodchikov (BPZ) 
power series representation \cite{BPZ}:\footnote{
In Eq.~(\ref{Bb}) symbols $V_{\Delta_i}(z_i)$ stand for Virasoro chiral vertex operators;
$\left[G_{c,\Delta}^{n}\right]^{IJ}$ is an inverse of the Gram matrix 
$\left[G_{c,\Delta}^{\,n}\right]_{IJ}=\langle\,\Delta_{I}^{n}\,|\,\Delta_{J}^{n}\,\rangle$
calculated in the basis $\left\lbrace|\,\Delta_{I}^{n}\,\rangle\right\rbrace$ 
of the subspace ${\cal V}_{c,\Delta}^{n}$
of the Verma module $\bigoplus\limits_{n=0}^{\infty}{\cal V}_{c,\Delta}^{n}$
with basis vectors labeled by partitions $I=(i_k \geq \ldots \geq i_1 \geq 1)$ 
with the length $n=i_k+\ldots+i_1=:|I|$.}
\begin{equation}\label{Bb}
	\label{block}
	{\cal F}\!\left(\,\cdot\,|\,x\,\right)=
	x^{\Delta-\Delta_{2}-\Delta_{1}}\left( 1 +
	\sum_{n=1}^\infty x^{n}\!\!
	\sum\limits_{n=|I|=|J|}
	\left\langle\,\Delta_4\,| V_{\Delta_3}(1) |\,\Delta_{I}^{n}\,\right\rangle
	\!\Big[G_{c,\Delta}^{n}\Big]^{IJ}
	\!\left\langle\,\Delta_{J}^{n}\,| V_{\Delta_2}(1) |\,\Delta_1\,\right\rangle\right).
\end{equation}
Mironov, Morozov and Shakirov (MMS) showed \cite{MMS} that ${\sf Z}(\,\cdot\,|{\bf z}_f)$ 
precisely reproduces the BPZ four-point block expansion.
Thus, there are two ways to compute the $b\To\infty$
asymptotic of ${\sf Z}(\,\cdot\,|{\bf z}_f)$. 
On the one hand, it's just a saddle point limit of the integral. 
On the other hand, it is the classical limit of the BPZ four-point block,
\begin{eqnarray*} 
{\cal F}\!\left(\Delta_i,\Delta,c\,|\,x\,\right)
\stackrel{b\to\infty}{\sim}{\rm e}^{b^2 f(\delta_i,\delta\,|\,x)}
&\;\;\Leftrightarrow\;\;&
f(\delta_i,\delta\,|\,x)\;=\;
\lim\limits_{b\to\infty}\frac{1}{b^2}
\log{\cal F}\!\left(\Delta_i,\Delta,c\,|\,x\,\right)\;=\;
\\[4pt]
&&\hspace{-100pt}\;=\;\left(\delta-\delta_1-\delta_2\right)\log x+
\frac{(\delta+\delta_2-\delta_1)(\delta+\delta_3-\delta_4)}{2\delta}\;x
+\ldots\;.
\end{eqnarray*}
This leads to the following result. 
\begin{list}{}{\itemindent=2mm \parsep=0mm \itemsep=0mm \topsep=0mm}
\item[1.] For
$\delta_i=\eta_i\left(\eta_i-1\right)$, $i=1,2,3$,
$\delta_4=\left(N_1+N_2+\eta_1+\eta_2+\eta_3\right)\left(N_1+N_2+\eta_1+\eta_2+\eta_3-1\right)$
and
$\delta=\left(N_1+\eta_1+\eta_2\right)\left(N_1+\eta_1+\eta_2-1\right)$
the classical four-point block on the sphere can be written in the following closed form \cite{PNPP}, i.e.,
\begin{eqnarray*}
f(\delta_i,\delta\,|\,x)&=&
-\,W(\,\cdot\,|{\bf z}_f,{\bf u}^{\sf c})
-\Big({\rm S}_{N_1}(2\eta_1,2\eta_2)+{\rm S}_{N_2}(2(\eta_1+\eta_2+N_1),2\eta_3)\Big)
\nonumber\\[5pt]
&+&2\eta_1\eta_2\log x+2\eta_2\eta_3\log(1-x),
\end{eqnarray*}
where $W(\,\cdot\,|{\bf z}_f,{\bf u}^{\sf c})$ is the critical value of the ``action'':
\begin{eqnarray*}\label{Wp}
	W(\,\cdot\,|{\bf z}_f,{\bf u})
	= -2\sum\limits_{\mu<\nu}\log(u_\nu-u_\mu)\nonumber
	-\sum\limits_{\mu=1}^{N_1+N_2}\left[2\eta_1\log u_{\mu}
	+2\eta_2\log\left(u_{\mu}-x\right)
	+2\eta_3\log\left(u_{\mu}-1\right)\right].
\end{eqnarray*}
\item[2.] Parameters ${\bf u}^{\sf c}=(u_{1}^{\sf c},\ldots,u_{N_1+N_2=N}^{\sf c})$
are solutions of the saddle point equations:
\begin{eqnarray*}\label{sadd}
\frac{\partial W(\,\cdot\,|{\bf z}_f,{\bf u})}{\partial u_{\mu}}&=&0\,
\;\Leftrightarrow\;
\frac{2\eta_1}{u_\mu}+\frac{2\eta_2}{u_\mu-x}+\frac{2\eta_3}{u_\mu-1}
+\sum\limits_{\nu\neq\mu}^{N}\frac{2}{u_\mu-u_\nu}=0,\\
&&\mu=1,\ldots, N=N_1+N_2.
\end{eqnarray*}
\end{list}

\noindent
The above statement can be generalized to the case of a multi-point spherical block \cite{PNPP}.
One sees that $W(\,\cdot\,|{\bf z}_f,{\bf u}^{\sf c})$ is a Virasoro analogue of the Coulomb energy 
$U({\bf z},{\bf u}^{\sf c})$ calculated in \cite{Sierra:1999mp}. 
This suggests that functions of this type are available as expansions of certain classical conformal blocks. 
On the other hand, the MMS techniques and the saddle point method provide a tool 
for summing expansions of classical blocks at least for certain values of the classical conformal weights.

MMS also proposed an integral representation of the one-point block on the torus (with modular parameter $\tau$)
and checked its consistency with the following $q$-series \cite{MMS2}: 
\begin{eqnarray*}\label{FtA}
{\cal F}^{\hat\Delta}_{c,\Delta}(q) &=& 
q^{\Delta-\frac{c}{24}}\sum\limits_{n=0}^{\infty}
{\cal F}^{\hat\Delta, n}_{c,\Delta}\,q^n,
\quad\quad q={\rm e}^{2\pi i \tau},
\\
{\cal F}^{\hat\Delta, n}_{c,\Delta}
&=&\sum\limits_{|I|=|J|=n}\left\langle\,\Delta_{I}^{n}\, | V_{\hat\Delta}(1) |\, \Delta_{J}^{n}\,\right\rangle
\Big[G_{c,\Delta}^{n}\Big]^{IJ}.\label{FtB}
\end{eqnarray*}
The MMS torus identity implies an analogous result to the one stated above, i.e.:
\begin{list}{}{\itemindent=2mm \parsep=0mm \itemsep=0mm \topsep=0mm}
\item[1.] For  $\hat\delta=N(N+1)$ and $\delta=\frac{1}{4}\left(a^2-1\right)$
the classical torus one-point block,
\begin{equation}\label{ft}
f^{\hat\delta}_{\delta}(q)=\left(\delta-\frac{1}{4}\right)\log q + \lim\limits_{b\to\infty}\frac{1}{b^2}\log
	\left[1+\sum\limits_{n=1}^{\infty}{\cal F}_{1+6Q^2,\Delta}^{\hat\Delta,n}\;q^n \right],
\end{equation}
can be written in the following finite form:
\begin{equation}\label{ft2}
f^{\hat\delta}_{\delta}(q)=\left(\delta-\frac{1}{4}\right)\log q-W\!(N,a,z^{c}_{1},\ldots,z^{c}_N),
\end{equation}
where
\begin{equation}
W(N,a,z_1,\ldots,z_N)=-\sum\limits_{r<s}2\log\theta_{*}(z_r-z_s)
+2N\sum\limits_{r=1}^{N}\log\theta_{*}(z_r)
-\sum\limits_{r=1}^{N}iaz_r,
\end{equation}
and  $\theta_{*}(z):=\sum_{n=0}^{\infty}(-1)^{n}q^{\frac{1}{2}n(n+1)}\sin\frac{(2n+1)z}{2}$.
\item[2.]
The parameters ${\bf z}^{\rm c}=(z^{c}_1,\ldots,z^{c}_N)$ are solutions of the saddle point equations
$\partial W/\partial z_r=0$, $r=1,\ldots,N$.
\end{list}

\noindent
The above result is new and will be discussed in detail in a separate paper. 
Here, we will just only announce that, based on this observation, one can connect the integral representation of the torus block
and its classical/saddle point limit with the Bethe ansatz approach to the elliptic Calogero-Moser (eCM) model.
The latter is a quantum many-body system with the $M$-particle Hamiltonian of the form \cite{takemura2000eigenstates}:
\begin{equation}\label{HM}
\hat{\sf H}^{\tau,\ell}_{M}:=-\frac{1}{2}\sum\limits_{i=1}^{M}\frac{\partial^2}{\partial z_{i}^{2}}+\ell(\ell+1)
\sum\limits_{1\leq i<j\leq M}\left(\wp(z_i-z_j,\tau)+2\eta\right),
\end{equation}
where $\ell\in\mathbf{Z}_{>0}$ is the coupling constant and $\wp(z,\tau)$ is the Weierstrass elliptic function.
In the 2-particle case the Hamiltonian (\ref{HM}) reads as follows
$\hat{\sf H}^{\tau,\ell}_{2}=-\frac{{\rm d}^2}{{\rm d}z^2}+\ell(\ell+1)\left(\wp(z,\tau)+2\eta\right)$, where $z=z_1-z_2$, 
and (cf.~\cite{takemura2000eigenstates}):
\begin{list}{}{\itemindent=2mm \parsep=0mm \itemsep=0mm \topsep=0mm}
\item[i.] the Bethe ansatz equations are given by
$\partial\Phi_{\tau}/\partial t_{i}=0$,
$i=1,\ldots,\ell$,
where
$$
\Phi_{\tau}(\ell,m_1,t_1,\ldots,t_\ell)={\rm e}^{i\pi\sum_{j=1}^{\ell}m_1t_j}
\prod\limits_{1\leq j\leq\ell}\theta(t_j)^{-2\ell}
\prod\limits_{1\leq i<j\leq\ell}\theta(t_i-t_j)^2,
$$
$$
\theta(x):=\frac{\hat\theta_{1}(x)}{\hat\theta_{1}'(0)}, 
\quad\quad
\hat\theta_{1}(x):=2q^{\frac{1}{8}}\sum_{n=0}^{\infty}(-1)^{n}q^{\frac{1}{2}n(n+1)}\sin((2n+1)\pi x);
$$
\item[ii.] the eigenfunction (Bethe vector) of the operator  $\hat{\sf H}^{\tau,\ell}_{2}$ is equal to
${\rm e}^{i\pi z}\theta(z-t_1)\ldots\theta(z-t_\ell)/\theta(z)^\ell$
up to a constant;
\item[iii.] the eigenvalue of the operator $\hat{\sf H}^{\tau,\ell}_{2}$ is equal to
\begin{equation}\label{S0}
{\rm const.}-2\pi i\partial_{\tau}S\!\left(t^{0}_{1},\ldots,t^{0}_{\ell};\tau\right),
\end{equation}
where $(t^{0}_{1},\ldots,t^{0}_{\ell})$ satisfy the Bethe ansatz equations and
$$
S\!\left(t_{1},\ldots,t_{\ell};\tau\right)= {\rm const.}\,2\sum_{i<j}\log\theta(t_i-t_j)-2\ell\sum_{i}\log\theta(t_i).
$$
\end{list}
The eigenvalue equation for the 2-particle eCM Hamiltonian is nothing but the famous Lam\'{e} equation, 
$\psi''(z)-\left[\,\kappa\,\wp(z) + {\sf B}\,\right]\psi=0$.
In CFT, one gets the latter from the classical limit of the null vector 
decoupling equation for the torus 2-point function with a light degenerate operator:
\begin{eqnarray*}\label{torus2point}
&&\left[\frac{1}{b^2}\,\frac{\partial^{2}}{\partial z^2}+
\left(2\Delta_+ \eta_1 + 2\eta_1 z \frac{\partial}{\partial z}\right) + \Delta_\beta
\left(\wp(z-w)+2\eta_1\right)\right.
\\[3pt]
&&\hspace{1cm}\left.+
\left(\zeta(z-w)+2\eta_1 w\right)
\frac{\partial}{\partial w}\right]\left\langle\, {\sf V}_{+}(z){\sf V}_{\beta}(w)\right\rangle_{\!\tau}
=
-\frac{2\pi i}{Z(\tau)}\;\frac{\partial}{\partial\tau}
\left[Z(\tau)\left\langle\,{\sf V}_{+}(z){\sf V}_{\beta}(w)\right\rangle_{\!\tau} \right].\nonumber
\end{eqnarray*}
In this way one can show that the Lam\'{e} eigenvalue ${\sf B}$ is given in terms of the classical torus block \cite{P1}:
\begin{equation}\label{B}
\frac{{\sf B}}{4\pi^2}=
q\frac{\partial}{\partial q}\,f^{\tilde\delta}_{\delta}(q)-\frac{\tilde\delta}{12}\,{\rm E}_{2}(\tau),
\quad\quad\tilde\delta=-\kappa,
\quad\quad q={\rm e}^{2\pi i \tau}.
\end{equation}
Combining (\ref{ft2}), (\ref{S0}) and (\ref{B}) one can expect that the critical value of 
$S\!\left(t_{1},\ldots,t_{\ell};\tau\right)$ in (\ref{S0}) is nothing but certain
classical torus one-point block. It would be interesting to investigate how it is in case of the $M$-particle operator 
(\ref{HM}).

\section{Discussion}
As a conclusion, we will share our thoughts on the possibility of using the classical 
limit of conformal blocks in further research on quantum many-body systems.

The Coulomb energy calculated in \cite{Sierra:1999mp} can be seen as the ``perturbed ${\rm su(2)}_{k}$ WZW classical block''.
It should be computable from the quantum block expansion.
The success of this idea would open the possibility of developing new techniques of finding energy spectra of quantum integrable
systems, which are alternative to the Bethe ansatz approach.  
Our preliminary calculations yield,  that also the classical block on the torus and the classical irregular block can be represented as critical values of the corresponding Coulomb gas integrals. 
The saddle point equations for the torus classical block are very similar to the Bethe ansatz equations for the eCM model.
We expect that in the case of the classical irregular block the corresponding integrable model will be the periodic Toda chain.
Finally, one can apply the KZ/BPZ correspondence \cite{Stoyanovsky} in the limit $c\To\infty$  to the integrable systems, 
integrability of which follows from the KZ equation (eg.,~the Richardson model). 
We suppose that in this way it will be possible to show that the classical Virasoro 
blocks determine energy spectra of these models.

There is one more formulation of the relationship between the Richardson model and CFT. 
This is an approach proposed by Sedrakyan and Galitski in \cite{SG} (see also \cite{SB}), 
which is close in spirit but technically different from the BCS/CFT correspondence 
discussed in \cite{Sierra:1999mp}. The authors of \cite{SG} 
asked whether there is a deformation of the SU(2) WZW model, such that the correlation 
functions of it are solutions of the modified KZ equation, which contains the integrals of motion of the Richardson model instead 
of just the Hamiltonians of the rational Gaudin model.
This deformed theory was identified in \cite{SG} as the boundary WZW model.
Authors of \cite{SG} have shown that the generalized KZ equation 
can be solved exactly using the so-called off-shell Bethe ansatz technique.
The solution of the latter can be given in an integral form. 
Analysis of this solution shows that this integral has a saddle 
point defined by the Richardson equations.
Here, the same question arises as before. If the saddle point value of the chiral correlation function represented by the 
appropriate integral solves the eigenvalue equations of Richardson conserved charges (which has not been shown until the end in 
\cite{SG}), then {\it does a certain ``classical block'' correspond to this  saddle point value?}
Moreover, one can ask directly about the limit $c\To\infty$ of the modified KZ equation. 
To understand what might happen here, we would like to use for this purpose
the correspondence between the BPZ and KZ equations \cite{Stoyanovsky}.
It turns out that the correspondence found by SG \cite{SG}
concerns a variety of dynamical systems that can be mapped on the boundary WZW model and solved exactly in many cases. 
Such an example is two-level laser with pumping and damping. Moreover, within the SG approach
one can study a nonequilibrium dynamics of various multi-level systems such as
models with time-dependent interaction strength, multi-level Landau-Zenner models and
some  many-body generalizations. An understanding of the nonequilibrium dynamics of quantum systems is
important in connection with quantum information problems and the idea of quantum computer (see refs.~in \cite{SB}).

The Coulomb gas integral (\ref{RCG}) of the Richardson model is known 
in the theory of random matrices as the so-called multi-Penner type $\beta$-ensemble with sources.
So, in parallel it is possible to use matrix models technology in this context.
Precisely, we would like to use a well-known calculation scheme within matrix models --- their  semiclassical 
('t Hooft) limit corresponding to the large-$c$ limit. This tool can be applied to investigate 
distributions of eigenvalues, i.e., the Richardson parameters (pair energies)
of the reduced BCS model. It would be interesting to compare this approach with the continuum limit of the 
Richardson equations, cf.~\cite{Roman2002}.
At least one reason is worth going in this direction. 
Gaudin proposed a continuum version of the Richardson equations. 
The assumption he made is that the solutions organize themselves into arcs $\Gamma_k$, 
$k=1,\ldots,{\rm K}$, which are symmetric under reflection on the real axis. 
For the ground state all the roots form a single arc ${\rm K}=1$. 
Still an open problem is \cite{Roman2002}:
``Study of solutions of Richardson equations with several arcs, i.e., ${\rm K}>1$ ... . ... they must describe very high excited states formed by separate condensates in interaction. This case may be relevant to systems such as arrays of superconducting grains or quantum dots. ..., the cases with ${\rm K} > 1$ seem to be related to the theory of hyperelliptic curves and higher genus Riemann surfaces, which may shed some light on this physical problem.''.  
The matrix models framework seems to be natural for these kinds of problems.

A fascinating open question concerning isolated quantum many-body systems 
is how they  evolve after a sudden perturbation or quench. For instance, 
in the paper \cite{WBHe} authors study a relaxation 
dynamics of the central spin model. 
Precisely, they analyze time evolutions of several 
quantities analytically and numerically.
It has been observed that the quantum dynamics of Gaudin magnets reveals a break-down of thermalization.
Methods used in the work \cite{WBHe} (the algebraic Bethe ansatz) do not go beyond those known from 
the Richardson solution and its implementation in CFT.  
Moreover, it is suggested in \cite{WBHe} to investigate scrambling and out-of-time-ordered 
correlators (OTOCs) for the Gaudin magnets. 
It should be stressed that OTOCs have recently been actively studied in the framework of CFT and 
these studies use the limit $c\To\infty$ of conformal blocks.
If it is possible to apply the large-$c$ limit of CFT to analyze OTOCs for the Gaudin magnets, 
it would be a very interesting research field for further exploration.


\section*{References}
\begin{thebibliography}{9}
\bibitem{ZZ}
Zamolodchikov A B, Zamolodchikov A B 1996 
Nucl.~Phys.~{\bf B} 477 577-605
	
\bibitem{PNPP} 
Pi\c{a}tek M R, Nazmitdinov R G, Puente A, Pietrykowski A R 2022
Classical conformal blocks, Coulomb gas integrals and Richardson-Gaudin models {\it J. High Energy Phys.} JHEP04(2022)098

\bibitem{R1} 
Richardson R W, 1963 
Phys.~Lett.~3 277-279 

\bibitem{RS}
Richardson R W, Sherman N 1964 
Nucl.~Phys.~52 221-238 

\bibitem{Gaudin1}
Gaudin M 1976 
J.~Phys.~37 1087-1098

\bibitem{Gaudin2}
Gaudin M 2014 {\it The Bethe Wavefunction} Cambridge University Press, Cambridge, translated by J.-S.~Caux

\bibitem{Cambiaggio:1997vz} 
Cambiaggio M C, Rivas A M F, Saraceno M 1997 
Nucl.~Phys.~{\bf A} 624 157-167

\bibitem{Sierra:1999mp}
Sierra G 2000 
Nucl.~Phys.~{\bf B} 572 517-534

\bibitem{BPZ}
Belavin A, Polyakov A, Zamolodchikov A 1984 
Nucl.~Phys.~{\bf B} 241 333-380

\bibitem{MMS}
Mironov A, Morozov A, Shakirov Sh 2010 
Int.~J.~Mod.~Phys.~{\bf A} 3173-3207

\bibitem{MMS2}
Mironov A, Morozov A, Shakirov Sh 2011 
J.~Phys.~{\bf A}:~Math.~Theor.~44 085401

\bibitem{takemura2000eigenstates}
Takemura K 2000 
Lett.~Math.~Phys.~53 181-194

\bibitem{P1}
Pi\c{a}tek M 2014 Classical torus conformal block, ${\cal N} = 2^*$ twisted superpotential and the accessory parameter of Lam\'{e} equation
{\it J. High Energy Phys.} JHEP03(2014)124

\bibitem{Stoyanovsky}
Stoyanovsky A V 2000 A relation between the Knizhnik-Zamolodchikov and Belavin-Polyakov-Zamolodchikov systems of partial differential equations ({\it Preprint} math-ph/0012013)

\bibitem{SG}
Sedrakyan T A, Galitski V  2010
Phys.~Rev.~{\bf B} 82, 214502

\bibitem{SB}
Sedrakyan T A, Babujian H M 2022
Quantum nonequilibrium dynamics from Knizhnik-Zamolodchikov equations {\it J. High Energy Phys.} JHEP04(2022)039

\bibitem{Roman2002}
Roman J M, Sierra G, Dukelsky J 2002
Nucl.~Phys.~{\bf B} 634 483-510.

\bibitem{WBHe}
He W B , Chesi S, Lin H Q, Guan X W 2022
Commun.~Theor.~Phys.~74 095102
\end{thebibliography}
\end{document}